\documentclass[twocolumn,           
               showpacs,            
               preprintnumbers,     
               aps,                 
               prd,                 
               a4paper,             
               superscriptaddress,  
               nofootinbib,         
               tightenlines,        
               floats,floatfix      
               ]{revtex4}

\usepackage{graphicx}
\usepackage{latexsym}
\usepackage{amsmath,amssymb}        
\usepackage[draft=false]{hyperref}

\begin{document}



\title{Cosmic microwave background multipole alignments in slab topologies}
\author{James G.~Cresswell}
\affiliation{Astronomy Centre, University of Sussex, Brighton BN1 9QH,
United Kingdom}
\affiliation{Institute of Cosmology and Gravitation, University of
Portsmouth, Portsmouth, PO1 2EG, United Kingdom}
\author{Andrew R.~Liddle}
\affiliation{Astronomy Centre, University of Sussex, Brighton BN1 9QH,
United Kingdom}
\author{Pia Mukherjee}
\affiliation{Astronomy Centre, University of Sussex, Brighton BN1 9QH,
United Kingdom}
\author{Alain Riazuelo}
\affiliation{CNRS, UMR 7095, Paris, F-75014 France; Universit\'e Pierre
et Marie Curie-Paris6, UMR 7095, Paris, F-75014 France}
\date{\today}
\pacs{98.80.-k,  98.70.Vc \hfill astro-ph/0512017}
\preprint{astro-ph/0512017}


\begin{abstract}
Several analyses of the microwave sky maps from the Wilkinson
Microwave Anisotropy Probe (WMAP) have drawn attention to alignments
amongst the low-order multipoles. Amongst the various possible
explanations, an effect of cosmic topology has been invoked by several
authors. We focus on an alignment of the first four multipoles ($\ell
= 2$ to $5$) found by Land and Magueijo (2005), and investigate the
distribution of their alignment statistic for a set of simulated cosmic 
microwave background
maps for cosmologies with slab-like topology. We find that this
topology does offer a modest increase in the probability of the
observed value, but that even for the smallest topology considered the
probability of the observed value remains below one percent.
\end{abstract}

\maketitle


\section{Introduction}

Several recent analyses of the WMAP satellite maps have pointed out an
unexpected degree of alignment between the low-order multipoles of the
cosmic microwave background (CMB) anisotropy
\cite{TDH,DTZH,Eriksen2003,Eriksen2004a,Hansen,Biel,LM,CHSS,BMRT}. Various
explanations have been put forward for these alignments, ranging from
statistical fluke or foreground contamination through to a genuinely
cosmological interpretation in terms of breakdown of statistical
isotropy. Such a breakdown would be a natural consequence of the
Universe possessing a non-trivial topology of characteristic scale
comparable to the observable Universe (for a selection of cosmic
topology review papers see Ref.~\cite{reviews}).

In this paper, we do not seek to address the interpretation of the
observational data, but rather aim to test whether or not slab-space
cosmic topologies give rise to the kind of alignments that are
tentatively reported to have been observed in the first-year WMAP
data. The observational indication is that there exists a preferred
direction for the low multipoles. For instance, Tegmark \textit{et
al.}~\cite{TDH} and de Oliveira-Costa \textit{et al.}~\cite{DTZH}
noted that the quadrupole and octupole were closely aligned with one
another, and approximately planar. Land and Magueijo \cite{LM}
(hereafter LM) sought the alignment for each multipole $\ell$ that
maximized the proportion of power contributed by a single $m$ mode,
and noted that the alignments of the first four multipoles were much
closer than would be expected under statistical isotropy. These
authors have all suggested that such alignments may be an indication
of a slab topology where only one dimension is compact (finite and
unbounded).

The principal aim of this paper is to determine whether the LM
alignment is a prediction of slab-space cosmic topology. We simulate
CMB maps for spatially-flat slab topologies, for different sizes of
the compact dimension, and derive the statistics of the alignments as
defined by LM. We find that the degree of alignment in the observed
data remain anomalous even in slab-space topologies.

\section{Cosmic Topology and the CMB}

If the Universe has a non-trivial topology, that would lead to a
breakdown of global isotropy. The spherical harmonic expansion
coefficients $a_{\ell m}$ of the observed map of CMB temperature
anisotropies would then no longer be uncorrelated random variables,
their correlation matrix having off-diagonal terms. Riazuelo
\textit{et al.}~\cite{Riazuelo04a,Riazuelo04b} found this correlation
matrix for many multiconnected spaces, by computing the eigenmodes of
the Laplacian with boundary conditions reflecting the particular
topology.

Let us briefly recall how this is done. The power spectrum
coefficients, the $C_\ell$, are computed via the formula
\begin{equation}
\label{topo1}
C_\ell \propto \int  \Theta_\ell^2 (k) P(k) k^2\;{\rm d} k \,,
\end{equation}
where $P(k)$ corresponds to the initial power spectrum of cosmological
perturbations of wavenumber $k$, and $\Theta_\ell (k)$ are transfer
functions defined in Ref.~\cite{huwhite95}.

In the case of Gaussian perturbations, the full statistical
information is encoded within the two-point correlation function. The
observable quantities correspond to the $a_{\ell m}$ coefficients of
the decomposition of the temperature field in spherical harmonics. The
quantity we are interested in is therefore the correlation matrix
$\left<a_{\ell m} a^*_{\ell' m'} \right>$. In a simply-connected
Universe, the cosmological principle implies that the cosmological
perturbations must be statistically isotropic. This in turn implies
that the above correlation matrix is necessarily diagonal:
$\left<a_{\ell m} a^*_{\ell' m'} \right> = \delta_{\ell \ell'}
\delta_{m m'} C_\ell$, where $C_\ell$ is defined above. In the case of
a multi-connected Universe, the Universe is no longer isotropic and
the correlation matrix has non-zero off-diagonal components.

In a spatially-flat multi-connected Universe, the eigenmodes of the
Laplacian $\Upsilon^k_s$ can be decomposed into the usual basis of
spherical harmonics and spherical Bessel functions as
\begin{equation}
\label{topo2}
\Upsilon^k_s = \xi^k_{s\ell m} \, j_\ell(kr) Y^m_\ell (\theta,
\varphi) \,,
\end{equation}
where the index $s$ distinguishes between modes with identical
wavenumber $k$.  With this decomposition, the coefficients of the
correlation matrix read
\begin{equation}
\label{topo3}
\left<a_{\ell m} a^*_{\ell' m'} \right> \propto \sum_{k, s}
\xi^k_{s\ell m} \xi^k_{s \ell' m'}{}^* \Theta_\ell (k) \Theta_{\ell'}
(k) P(k) \,.
\end{equation}
Once the $\xi^k_{s\ell m}$ coefficients are known, all the statistical
information about a given topology can be computed using an existing
CMB code. This method therefore naturally takes into account all the
contributions (Sachs--Wolfe, Doppler and Integrated Sachs--Wolfe
(ISW)) to the temperature anisotropies, unlike the analytical formula
of Ref.~\cite{DTZH} which performs an estimate of the Sachs--Wolfe
term only in the large-wavelength limit.

Should one take into account the Sachs--Wolfe contribution only, then
the CMB maps would exhibit sets of pairs of circles whose temperature
pattern would perfectly match. These correlated temperature patterns
arise from the fact that we see two copies of the same region along
different lines of sight. The Doppler and ISW contributions reduce
this correlation because the Doppler term depends on the direction in
which the electron velocity field is observed, and the ISW effect
depends on the photon history along the line of sight \cite{CDS}. The
simulated maps neglect reionization.  As with the ISW effect,
reionization tends to blur the topological signature in the
correlation matrix, so that our simulated maps exhibit a stronger
departure from statistical isotropy than more realistic maps.

Following LM's suggestion, we restrict our study to slab spaces. Our
computational method requires all directions to be finite, so we chose
rectangular tori with dimensions of the form $15 \times 15 \times X$,
labelled T$[15,15,X]$ in the notation of Kunz {\it et
al.}~\cite{Kunz2005a}, where $X = 1, 2, ..., 15$ and the sizes are in
Hubble radius units.  Recall that the distance to the last-scattering
region today is around $3.1$ Hubble radii in a flat $\Lambda$CDM model
with $\Omega_\Lambda \simeq 0.7$.  The dimensions of size 15 Hubble
units are essentially infinite, which we checked by comparing the
correlation matrix of T$[15,15,15]$ to that of a standard simply
connected, infinite universe.  Hence T$[15,15,X]$ is a
computationally-favourable approximation to a slab space (that is, a
space with only one compact dimension).  Another reason to consider
slab spaces is that matched circles searches have so far given
negative results \cite{starkman}, so that it seems likely that only
topologies exhibiting a small number of circles, such as slab spaces,
are compatible with the data.

\section{Results}

\subsection{The Land--Magueijo statistic and its observed value}

LM devised a statistic to study the alignments of multipoles. For each
multipole, they found the orientation of the coordinate axes which
maximized the concentration of the multipole power into a single $m$,
defining
\begin{equation}
r_\ell = \max_{m,{\bf n}} \frac{C_{\ell m}}{(2\ell+1)C_\ell} \,,
\label{rl statistic}
\end{equation}
where ${\bf n}$ is the coordinate axis orientation, $C_\ell$ the usual
power spectrum, and $C_{\ell m}$ measures the power at a single $m$
value, defined as $C_{\ell 0}=\left|a_{\ell 0}\right|^2$ and $C_{\ell
m} = 2\left|a_{\ell m}\right|^2$ for $m>0$. The vector ${\bf n}_\ell$
is defined as the one which provides the maximum value of the
statistic $r_{\ell}$ for each $\ell$.

They noticed a strong alignment of the ${\bf n}_\ell$ of the lowest
four multipoles, $\ell = 2$,...,$5$ and quantified this by defining
the mean angle between the six different pairs of alignments. As the
orientation vectors are headless (the same results are achieved by
interchanging ${\bf n}_\ell$ with $-{\bf n}_\ell$), one must choose
the angle $\theta_{ij}$ which is less than 90 degrees. The average
alignment angle is then
\begin{equation}
\hat{\theta} = \mathrm{mean} \left( \theta_{ij} \right), \quad
i,j=2,3,4,5 \; \mbox{with} \; i \neq j \,.
\label{single mean theta}
\end{equation}

They evaluated this alignment angle for the Tegmark, de
Oliveira-Costa, Hamilton (TOH) cleaned and Wiener filtered maps of
Ref.~\cite{TDH} and found the values of $22.4^\circ$ and $22.3^\circ$
respectively.\footnote{Their paper quotes `of order of 20 degrees'; we
thank Kate Land for providing the actual value.} We have confirmed
this result using the code described below.

For a Gaussian map, the predicted value of $\hat{\theta}$ is one
radian. This arises as follows.\footnote{Thanks to Kate Land for
providing this argument, not given in their original paper.}
Remembering that the alignment direction is a headless vector, the
average separation between two such vectors can be found by rotating
coordinates so that one is at the north pole. The other then has one
end uniformly distributed in the northern hemisphere, so the average
angle is given by the average distance of a point in the northern
hemisphere from the north pole, which is one radian.

Land and Magueijo found that the low observed value arose in only 5
out of 5000 simulated Gaussian maps, a result which we confirm
below. On the face of it, this strongly excludes statistical isotropy,
but one does need to bear in mind the strong {\it a posteriori}
selection of their statistic; for instance the signal would be much
weaker if the average angle included alignment with even just the
sixth multipole.

It is also found that because the statistic involves maximization, it
can be highly sensitive to small changes in the map, because multipole
concentrations can have near double maxima for different $m$, giving
completely different alignment directions.  For example, while in the
result above using the TOH maps, the quadrupole alignment was one
which maximized the $m=2$ multipole, with ${\bf n}_2$ in the direction
$(b,l)\approx(60,-100)$, we find (as did LM) that for the Lagrange
internal linear combination map (LILC) produced by Eriksen \textit{et
al.}~\cite{Eriksen2004a} a completely different orientation is
selected for the quadrupole, this time maximizing the $m=0$ multipole
(in the TOH map this maximum is just slightly less than the one
chosen), which significantly increases $\hat{\theta}$ to $55.2^\circ$.
Similarly, in the case of the internal linear combination map (ILC)
produced by the WMAP team \cite{Bennett03}, while the preferred axis
of the quadrupole is similar to that in the TOH maps, this time it is
the direction of the octupole that is the cause of a discrepancy that
increases $\hat{\theta}$. We note this as a caveat; in both these
cases of near double maxima, three of four maps agree as regards the
direction of the preferred axis for the concerned multipole. The maps
differ in the details of how foregrounds are removed from them.

Again, we do not here seek to address data-related aspects such as the
likelihood of finding the observed $\hat{\theta}$ given the
uncertainties in foreground subtraction, the sky-cut, cosmic variance
etc. While these issues are important to consider when determining
whether the claimed detection of a breakdown of isotropy is
significant, papers that reported the detections in the first place
have delved into such questions to some extent, and there is only
limited progress one can make with an {\it a posteriori}
detection. Aspects of foreground contamination have been studied in
Refs.~\cite{CHSS,Eriksen2004a,Biel2,Vale,MC}. For instance, Copi
\textit{et al.}~\cite{CHSS} considered the nature of known foregrounds
without finding any clear connection to low-$\ell$ alignments, at
least as regards the LM kind.

Despite the above-mentioned caveats, the value of $\hat{\theta}$ found
by LM is very low, and they speculate that it may be a signature of
slab topology, already invoked in Refs.~\cite{TDH,DTZH} as a possible
explanation of the quadrupole and octupole planarity. Our aim is to
test this suggestion by evaluating the distribution of the
$\hat{\theta}$ statistic for simulated slab topology maps, and
checking whether, relative to the case of a trivial topology, such
topologies can better explain a low $\hat{\theta}$.

\subsection{The Land--Magueijo statistic for simulated maps of slab
topology}

Given the full ensemble average correlation matrix $\langle a_{\ell
m}a^*_{\ell' m'} \rangle$ of the spherical harmonic coefficients for
slab space topologies of form T$[15,15,X]$, we create corresponding
random realizations of the $a_{\ell m}$. These are then rotated
through a two-dimensional grid of galactic angles $b$ and $l$ (we used
spacings of one degree in each). The $a_{\ell m}$ transform under
rotations as (see e.g.~Ref.~\cite{CDEW})
\begin{equation}
a_{\ell m}=
\sum_{m^{\prime}=\,-\ell}^{\ell} a_{\ell m^{\prime}} e^{-im\alpha} 
d_{m\,m^{\prime}}^{\, \ell} (\beta) e^{-im^{\prime}\gamma} \,,
\label{alm rotation}
\end{equation}
where ($\alpha,\beta,\gamma$) are the Euler angles corresponding to
the rotation, and $d_{mm'}^\ell(\alpha,\beta,\gamma)$ is part of the
representation of the Wigner rotation matrix \cite{CDEW}. Rotating the
$a_{\ell m}$ over the entire grid, the rotation $\mathbf{n}_\ell$ (and
associated $m$) that maximized $r_\ell$ was recorded, for each
$\ell$. Thus $\hat{\theta}$ was found for each simulation. This was
repeated for an ensemble of universes so that the distribution of
$\hat{\theta}$ was built up.  The ensemble average value
$\langle{\hat{\theta}}\rangle$ was also found. This was then repeated
for different dimensions $X$ of the rectangular toroid T$[15,15,X]$.

\begin{figure}[t]
\includegraphics[width=7.5cm]{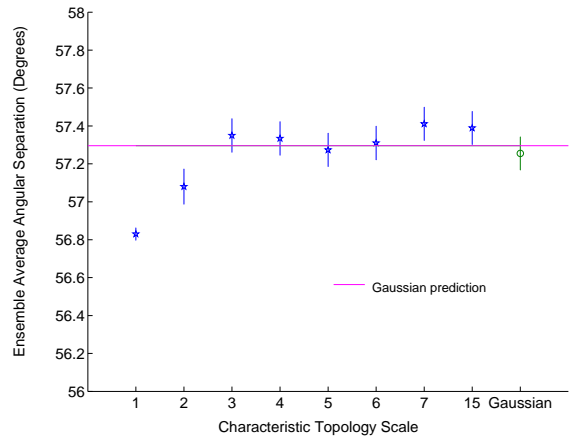}
\caption{The ensemble average alignment $\langle{\hat{\theta}}\rangle$
and its standard error for the T$[15,15,X]$ topologies, as a function
of topology scale $X$. The horizontal line is the Gaussian
prediction.}
\label{fig:values}
\end{figure}

Figure~\ref{fig:values} shows the ensemble average values
$\langle{\hat{\theta}}\rangle$, together with their standard error,
against $X$. For Gaussian maps, the expected result of one radian is
accurately recovered. We further see that the large-scale topology
maps behave essentially as Gaussian maps.  Only for $X < 3$ do we
begin to see an effect of topology, with the ensemble average reducing
slightly with respect to the Gaussian result. The shift is very small
as compared to the observed value of $\hat{\theta}$.

\begin{figure*}[t]
\includegraphics[width=7.5cm]{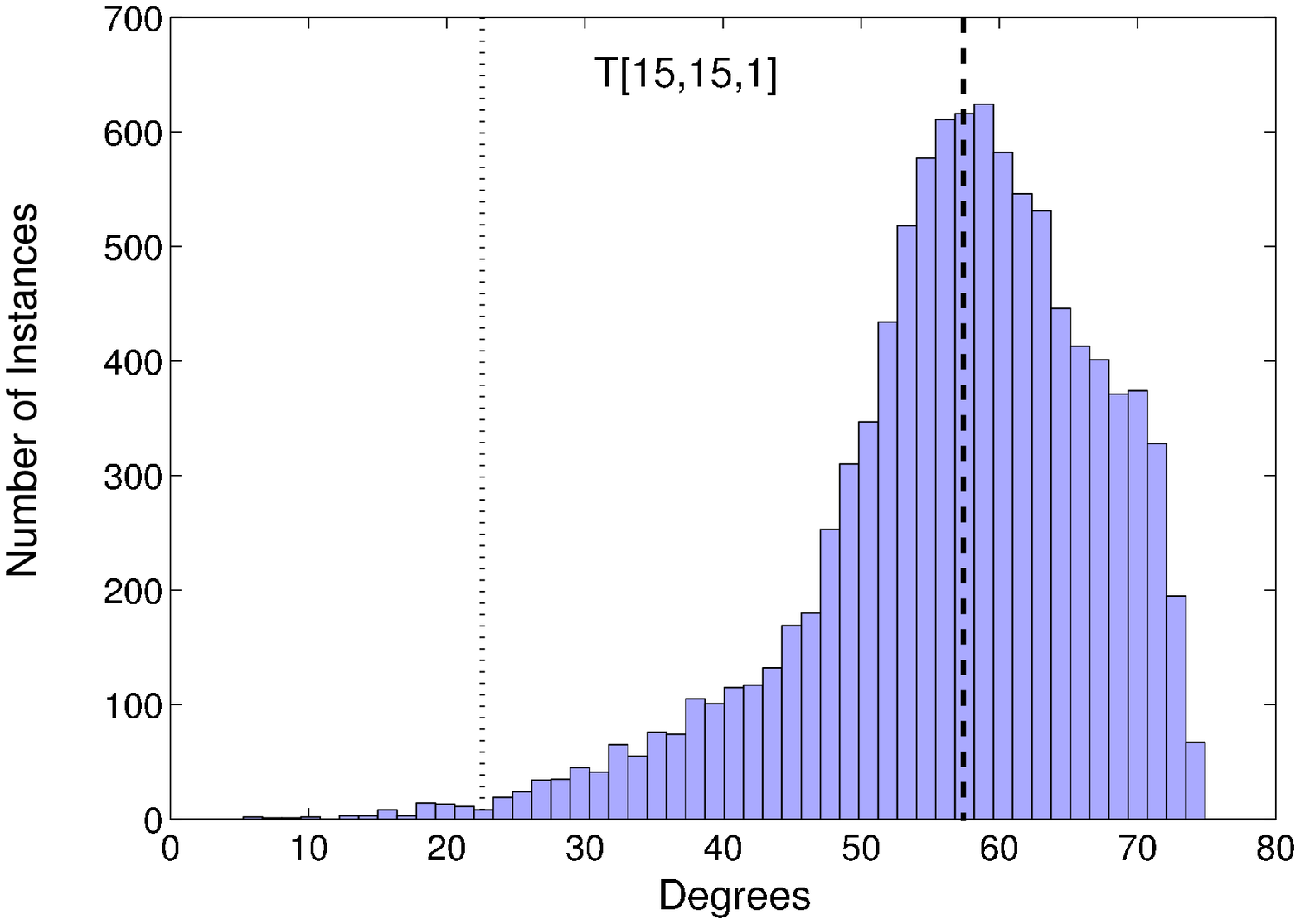} \hspace*{1cm}
\includegraphics[width=7.5cm]{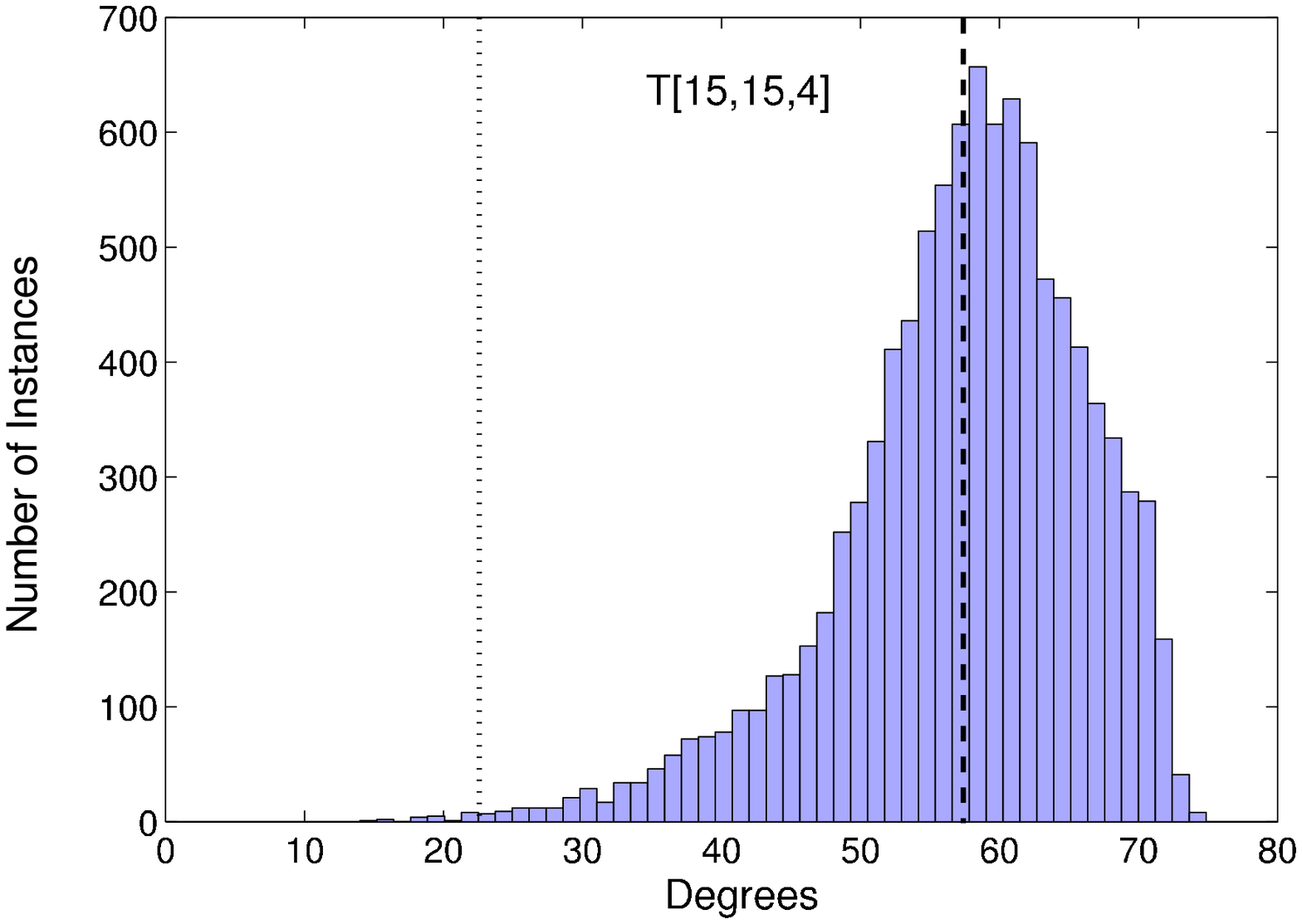}\\
\includegraphics[width=7.5cm]{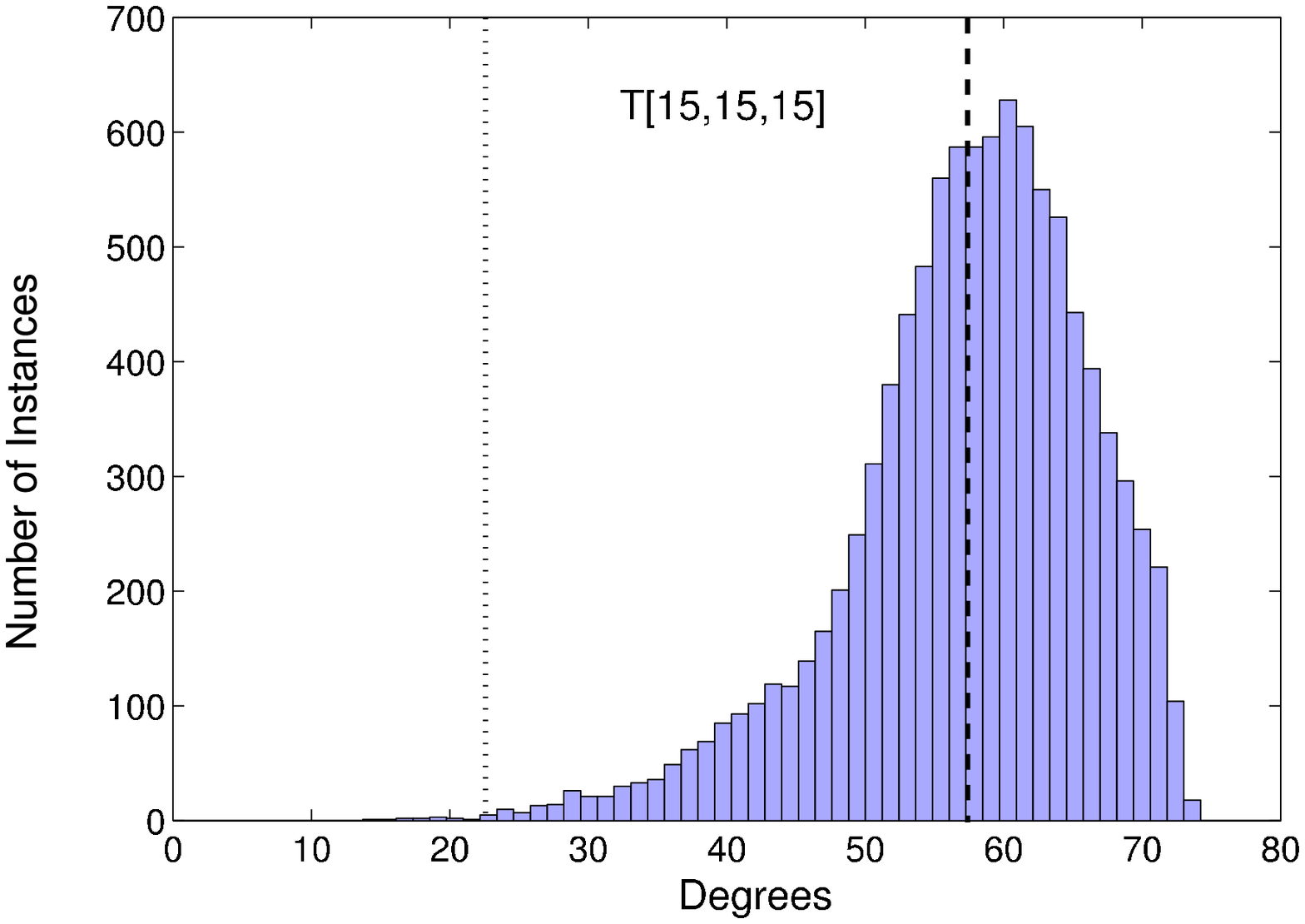} \hspace*{1cm}
\includegraphics[width=7.5cm]{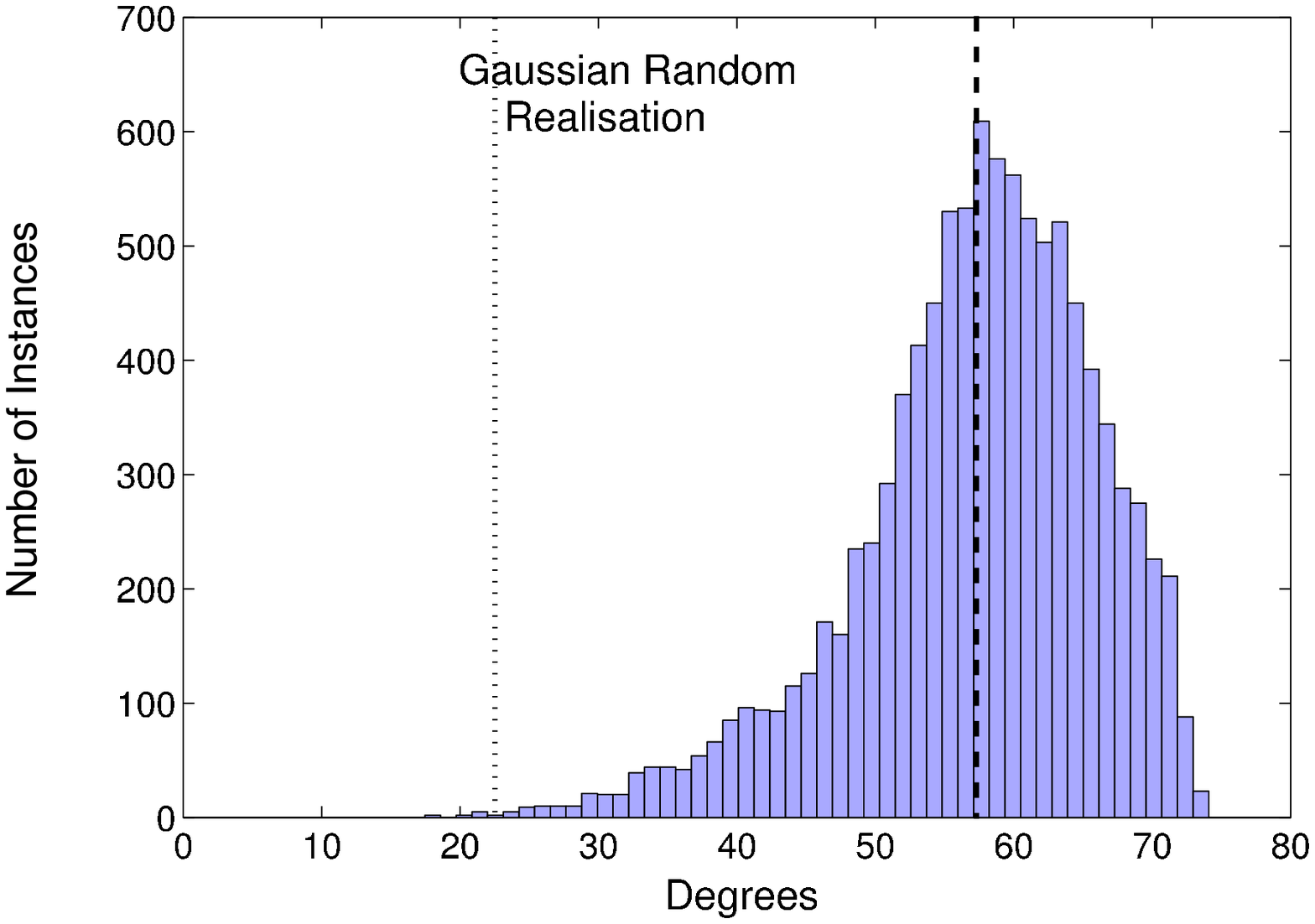}
\caption{The full distribution of $\hat{\theta}$ for three
T$[15,15,X]$ topologies, $X=15,4,1$, and for a Gaussian random
realization. The dashed line shows the mean value and the dotted line
the observed value. Note the extended low-end tail of the T[15,15,1]
distribution, corresponding to the increased probability of this
topology being able to produce the observed result.}
\label{fig:histo}
\end{figure*}

Although the shift of the ensemble mean is small, topology might
nevertheless help explain the observation if it alters the
distribution of $\hat{\theta}$ for small angles.
Figure~\ref{fig:histo} shows the full distribution of mean angle
$\hat{\theta}$ obtained from 10,000 realizations of the corresponding
topologies. Such a large number of simulations was used in order to
trace the tails of the distribution accurately. The value of one
radian, which is the expected mean angle for a Gaussian random
realization, is shown with a thick dashed line, and the observed value
is shown as a dotted line. While the distribution of $\hat{\theta}$
extends a little towards smaller values as the size of the smaller
dimension of the toroid decreases, the observed value remains
significantly low even for the smallest dimension considered.

\begin{figure}
\includegraphics[width=7.5cm]{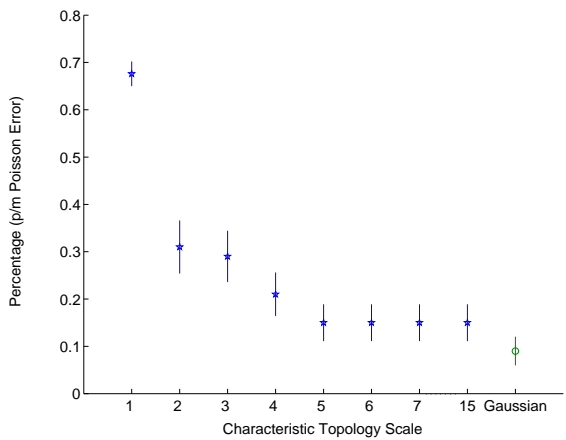}
\caption{The fraction of realizations of a given T$[15,15,X]$ topology
which provide a result at least as low as the observed result
($\hat{\theta}=22.5^{\circ}$), shown as a function of the
characteristic topology scale $X$.}
\label{fig:likelihood}
\end{figure}

We quantify this further in Figure~\ref{fig:likelihood}, which shows
the fraction of the ensemble giving a value at least as low as the
observed one. The uncertainties are estimated using the Poisson error
on the number of such ensemble members. We confirm the result of LM
that approximately 0.1\% of Gaussian skies give the observed value or
lower. For small topology scales we see an enhancement in this
fraction, but even for the smallest topology considered, the
probability remains below one percent.  We conclude that the observed
alignment is not predicted by slab-space topologies.

We have only asked whether the topologies within our set can explain
the result found by LM. This is not achieved even by the smallest
topologies we consider. However we note further that such small
topologies are almost certainly already excluded by other observations
\cite{OSS}. For instance, a harmonic space analysis of precisely these
same simulated topologies \cite{Kunz2005a} suggests that $X<3$ is
excluded by comparison to WMAP data (see also Ref.~\cite{PK}). Such
topologies are also constrained by the null results of matched circles
tests for such topologies \cite{CDS,starkman}, corresponding to a
limit of about $X \sim 5$. [The very tentative indications of
dodecahedral topology found in Ref.~\cite{Lum} are interesting but of
no significance for the topologies we consider here.] We have found
that such topologies also do not receive any real support even from
the observed alignments of the type discussed here.

\subsection{The Land--Magueijo statistic for ILC simulations}

Stepping somewhat aside from the main drive of this paper, we have
made a small study of foreground effects by finding the distribution
of $\hat{\theta}$ that results from the 10000 ILC simulations provided
by Eriksen \textit{et al.}~\cite{Eriksen2005}.  These simulations
additionally (over an assumed Gaussian CMB sky) contain the residual
level of foregrounds that can be expected from the ILC method of
foreground subtraction. The distribution of $\hat{\theta}$ that
results is much broader and flatter towards smaller angles. That is
expected because the simulations contain residuals of the galaxy, so
that a preferred axis going through the galactic poles or thereabouts
will be expected for the lower $\ell$ and hence $\hat{\theta}$ will be
smaller.

This result does not, however, explain the Land--Magueijo result, as
the alignment they find (which is identified in the TOH maps) is not
directed towards the galactic poles.  It seems that if there were a
preferred axis in a Gaussian CMB sky contaminated by galactic
residuals, then it would be in the direction of the poles.  This
agrees with the conclusion of Copi \textit{et al.}~\cite{CHSS} that
the alignments seen do not correlate with known foregrounds.  Also,
note that the real ILC maps do not show the same alignment direction
effect, the effect being spoilt due to the previously-mentioned
presence of a near double maxima in one of the four multipoles
considered.

To summarize, foreground contamination could explain a small
$\hat{\theta}$ about the poles, but the observed orientation is not
explained.  Further, the alignment effect is detected in maps using a
different foreground cleaning method \cite{TDH}. Hence this may be a
case where a curious real feature is being obscured by the presence of
galactic contamination (and cosmic variance). Interestingly, this
point has already be made for related alignment statistics by Slosar
and Seljak \cite{SS}.

~ 

\section{Conclusions}

Our main results are as follows. We have confirmed the observed value
of $\hat{\theta}$ found by Land and Magueijo, while noting that it is
quite dependent on the choice of maps used. We have also confirmed
their result that Gaussian skies have only about 0.1\% chance of
finding a value as low as that observed in the TOH maps. By analyzing
a set of slab-topology maps, we have found that there is a
slightly-enhanced probability of such a low value being obtained, but
in absolute terms it remains extremely unlikely. We conclude that slab
topology is not the explanation for the multipole alignment found by
Land and Magueijo. The resolution must lie elsewhere, perhaps in other
topologies, or instead in other cosmological assumptions, or in
foreground or instrumental noise.


\begin{acknowledgments}

J.G.C., A.R.L.~and P.M.~were supported by PPARC. We thank Martin Kunz
and Kate Land for helpful discussions, and Patrick Dineen for
providing the multipole rotation code used in Ref.~\cite{CDEW}.  Many
of our results were generated using the HEALPix package \cite{HEAL}.
\end{acknowledgments}


\end{document}